\documentclass{article}
\usepackage{amsmath}
\usepackage{amsfonts}
\usepackage{amssymb}
\usepackage{graphicx}

\begin{document}

\title{Expanding Self-Graviting Dust Ball }
\author{L.V.Verozub}
\begin{abstract}
In this paper we consider dynamics of a self - graviting matter in flat space
- time starting from the gravitation equations proposed in the paper
\cite{Verozub1}. It is shown that only "asymtotical" singularity in the
infinitely remote past occurs in the model. In connection with evidence that
the deceleration parameter $q_{0}$ in cosmology is negative (\cite{Riess},
\cite{Perlmutter}) it is shown that a supermassive dust ball expands with
acceleration from zero up to some radius that depends on the model parameters .
\end{abstract}
\maketitle

\section{ Introduction.}

Thirring \cite{Thirring} proposed that gravitation can be described as a
tensor field $\psi_{\alpha\beta}(x)$ of spin two in Pseudo-Euclidean
space-time $E_{4}$ where the Lagrangian action describing the motion of test
particles in a given field is of the form
\begin{equation}
L=-m_{p}c\left[ g_{\alpha\beta}(\psi)\dot{x}^{\alpha}\dot{x}^{\beta}\right]
^{1/2}\;.
\end{equation}

In this equation $g_{\alpha\beta}$ is a tensor function of $\psi_{\alpha\beta
}$, $m_{p}$ is the mass of the particle , $c$ is the speed of light and
$\dot{x}^{\alpha}=dx^{\alpha}/dt$ .

A theory based on that action must be invariant under the gauge
transformations $\psi_{\alpha\beta}\longrightarrow\bar{\psi}_{\alpha\beta}$
that are a consequence of the existence of ''extra'' components of the tensor
$\psi_{\alpha\beta}$. The transformations $\psi_{\alpha\beta}\longrightarrow
\bar{\psi}_{\alpha\beta}$ give rise to some transformations $g_{\alpha\beta}$
$\longrightarrow$ $\bar{g}_{\alpha\beta}$ . Therefore , the field equations
for $g_{\alpha\beta}(x)$ and equations of the motion of the test particle must
be invariant under these transformations of the tensor $g_{\alpha\beta}$ .
Equations of gravitation that are invariant with respect to arbitrary gauge
transformations were proposed in the paper \cite{Verozub1}. These equations
are of the form%

\begin{equation}
B_{\alpha\beta;\gamma}^{\gamma}-B_{\alpha\delta}^{\epsilon}B_{\beta\epsilon
}^{\delta}=0\label{myeqs}%
\end{equation}

The equations are vacuum bimetric equations for the tensor
\begin{equation}
B_{\alpha\beta}^{\gamma}=\Pi_{\alpha\beta}^{\gamma}-\overset{\circ}{\Pi
}_{\alpha\beta}^{\gamma}\label{tensB}%
\end{equation}
(Greek indices run from 0 to 3) , where
\begin{equation}
\Pi_{\alpha\beta}^{\gamma}= \Gamma_{\alpha\beta}^{\gamma}-(n+1)^{-1} \left[
\delta_{\alpha}^{\gamma}\Gamma_{\epsilon\beta}^{\epsilon}-\delta_{\beta
}^{\gamma}\Gamma_{\epsilon\alpha}^{\epsilon}]\right]  ,\label{Thomases}%
\end{equation}%

\begin{equation}
\overset{\circ}{\Pi}_{\alpha\beta}^{\gamma}=\overset{\circ}{\Gamma}%
_{\alpha\beta}^{\gamma}-(n+1)^{-1}\left[  \delta_{\alpha}^{\gamma}%
\overset{\circ}{\Gamma}_{\epsilon\beta}^{\epsilon}-\delta_{\beta}^{\gamma
}\overset{\circ}{\Gamma}_{\epsilon\alpha}^{\epsilon}\right]
,\label{Thomases0}%
\end{equation}

$\overset{\circ}{\Gamma}_{\alpha\beta}^{y}$ are the Christoffel symbols of the
pseudo-Euclidean space-time $E_{4}$ whose fundamental tensor is $\eta
_{\alpha\beta}$ , $\Gamma_{\alpha\beta}^{\gamma}$ are the Christoffel symbols
of the Riemannian space-time $V_{4}$ of the dimension $n=4$ , whose
fundamental tensor is $g_{\alpha\beta}$ . The semi-colon in eqs. (\ref{myeqs})
denotes the covariant differentiation in $E_{4}$.

The peculiarity of eqs.(\ref{myeqs}) is that they are invariant under
arbitrary transformations of the tensor $g_{\alpha\beta}$ retaining invariant
the equations of motion of a test particle , i.e. geodesics in $V_{4}$ . In
other words, the equations are geodesic-invariant . Thus , the tensor field
$g_{\alpha\beta}$ is defined up to geodesic mappings of space-time $V_{4}$ (In
the analogous way as the potential $A_{\alpha}$ in electrodynamics is
determined up to gauge transformations). A physical sense has only geodesic
invariant values. The simplest object of that kind is the object
$B_{\alpha\beta}^{\gamma}$ which can be named the strength tensor of
gravitation field . The coordinate system is defined by the used measurement
instruments and is a given.

\section{Evolution of an Expanding Dust-Ball}

Consider in flat space-time an expanding sphericaly symmetric homogeneous dust
- ball with the sizes of the observed Universe.

The motion of the specks of dust of the sphere $R$ are described by the
following Lagrangian \cite{Verozub1}
\begin{equation}
L=-m_{p}c\left[  c^{2}C-A\dot{r}^{2}-f^{2}(\dot{\varphi} ^{2}\sin^{2}%
\theta+\dot{\theta}^{2})\right]  ^{1/2},\label{lagrangian2}%
\end{equation}
where
\[
A=r^{4}/f^{4}(1-r_{g}/f),\ \ \ C=1-r_{g}/f,
\]%

\[
f=(r_{g}^{3}+r^{3})^{1/3},\ \ \ r_{g}=2GM/c^{2}
\]
, $\dot r = dr/dt$, $\varphi= d\varphi/dt$, $\dot\theta=d\theta/dt$, $M$ is
the mass of the dust-ball, $G$ is the gravitational constant , $A$ and $C$ are
the functions of r .

The differential equations of the particles at the radial motion of the sphere
surface with the radius $R$ are given by
\begin{equation}
\dot{R}^{2} = \frac{c^{2}C}{A} \left[  1- \frac{C}{\bar E^{2}} \right]
,\label{eq1_motion}%
\end{equation}

where , $\dot R =dR/dt$, $\bar E =E/m_{p}c$ and $E$ is the energy of the
specks of dust,$C$ and $A$ are the functions of $R$.

Setting $\dot R =0$ in eq. (\ref{eq1_motion}) we obtain
\begin{equation}
\bar E^{2}=\mathcal{N}(R),\label{efpot}%
\end{equation}
where
\[
\mathcal{N}(R)=1-\frac{r_{g}}{f}
\]

is the effective potential of the gravitational field.

We assume here, for definiteness, that at the moment the radial velocity is
$V_{0}=H_{0}R_{0}$ where $H_{0}=0.25\cdot10^{-17}s^{-1}$ is the Habble
constant and $R_{0}=3\cdot10^{27}cm$ is the radius .

The function $\mathcal{N}(R)$ is the effective potential of the sphericaly
symmetric gravitational field in the theory under consideration. Fig. 1 shows
the function $\mathcal{N}(R)$ for the density $\rho=10^{-28}gm/cm^{3}$ . The
constant $\bar{E}^{2}$ is equal to $0.60$ .%

\begin{figure}
[h]
\begin{center}
\includegraphics[
height=1.9112in,
width=3.0943in
]%
{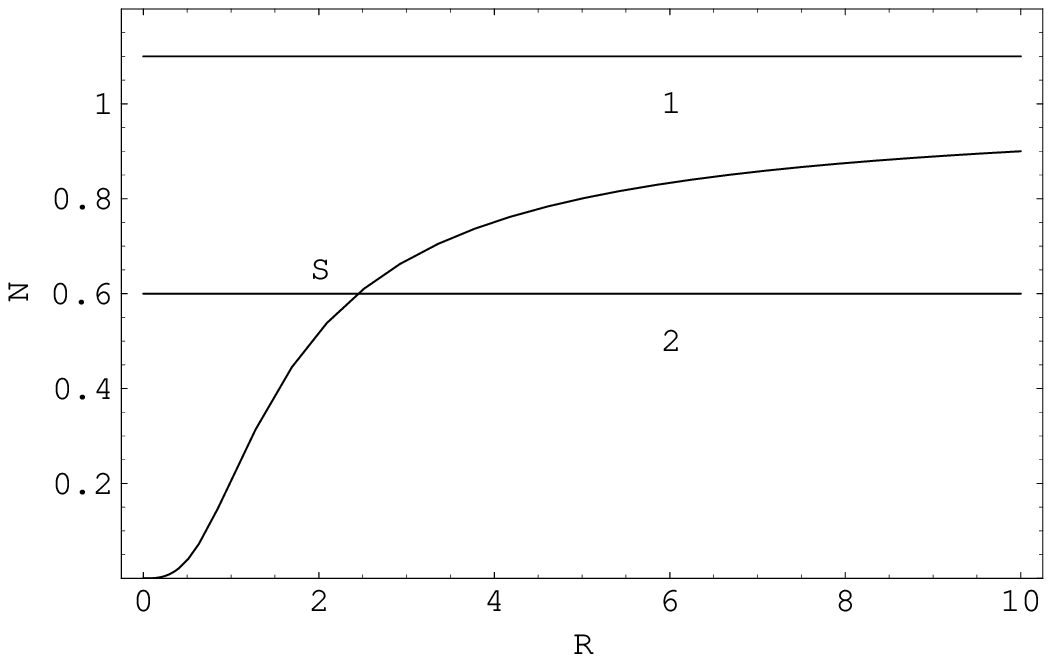}%
\end{center}
\end{figure}

Straight lines $\bar E^{2}=Const$ show possible types of the particles orbits
(the expansion scenarios) :

1. The orbits with .$\bar E^{2}\geq1$ (Straight line 1). The expansion begins
at $R=0$ and continues to the infinity.

2. The orbits with $\bar E^{2}\leq1$. A) The expansion begins at $R=0$ and
continues up to the point of the line crossing with the curves $\mathcal{N}%
(R)$. (The point $S$ in the figure) . B) The expansion begins at the point $S
$ and continues to the infinity. (Evidently, the latest possibilty is not realized).

Fig.2 shows the dependence of the sphere radius on the time (curve 1) and the
same function in the Newtonian theory (curve 2)

\vspace{0.5cm}%
\begin{figure}
[h]
\begin{center}
\includegraphics[
height=1.9121in,
width=3.103in
]%
{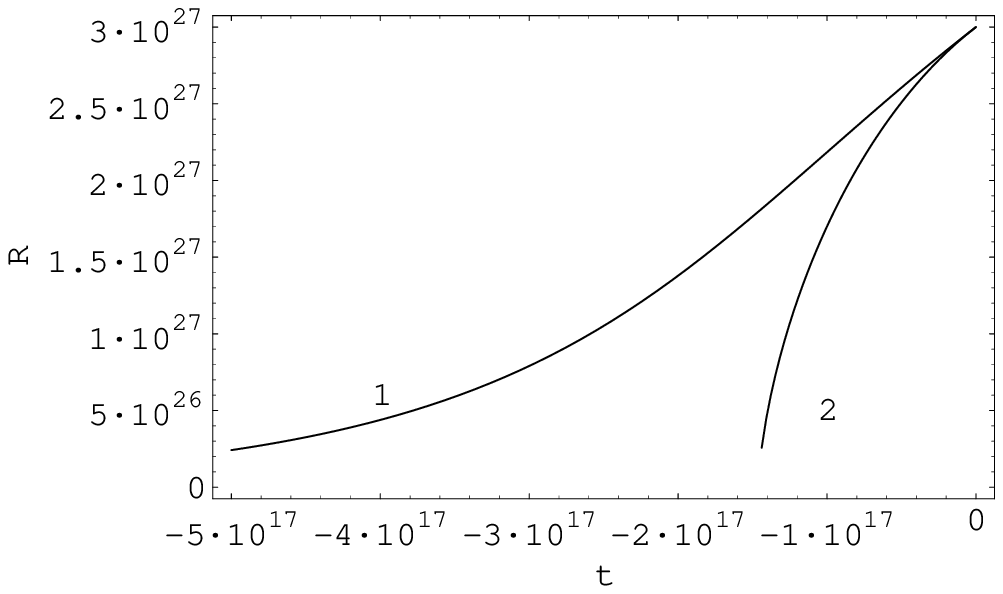}%
\end{center}
\end{figure}

The time
\begin{equation}
T=\int_{R_{in}}^{R_{0}}\dot{r}^{-1}dr\label{time_univ}%
\end{equation}
of the expansion tends to zero if $R_{in}$ tends to zero .Therefore only
''asymptotical'' singularity in the infinitely remote past ocuurs in the model.

The expansion will continue as far as the point $S$ in Fig. 1 where the radius
$R=0.42\cdot10^{28}cm$. Because of the radial velocity decreasing it will
happen in $0.12\cdot10^{19}s$ .

Analysis of the recent observations data gives evidence that the deceleration
parameter $q_{o}=-\ddot{a}(t)$ $a(t)/\overset{.}{a}(t)$ ( $a$ is the scale
factor) is negative at the moment (\cite{Riess} , \cite{Perlmutter}) . It
means that $\ddot{a} >0$ i.e. the expansion is accompanied with acceleration,
while according to classical insights the gravity force must retard the
expansion. Let us show that from the viewpoint of our gravitation equations
(\cite{Verozub1}) the dust ball expands with acceleration from $R=0$ up to
some radius depending on the physical parameters of the model.

Fig. 3 shows the plot of the velocity $V=\dot{R}$ and the acceleration
$dV/dt=(dV/dR)V$ from $R=0$ up to the end .%

\begin{figure}
[h]
\begin{center}
\includegraphics[
height=1.9121in,
width=3.103in
]%
{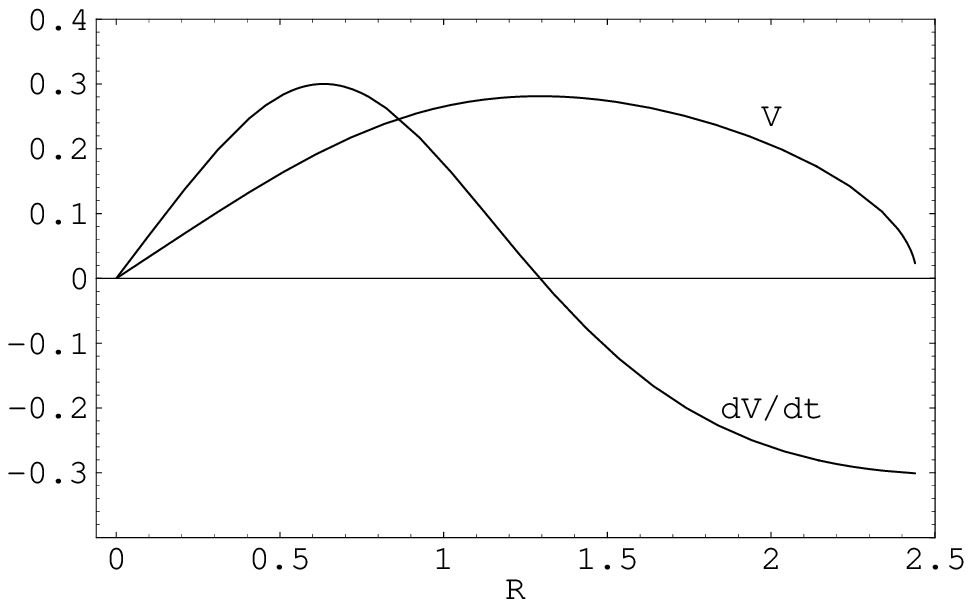}%
\end{center}
\end{figure}

It follows from the plot that the acceleration is negative at $R/r_{g}>1.3$ .
However, according to the properties of gravity force following from
\cite{Verozub1} it is a positive value at $R/r_{g}<1.3$ . At the density
$\rho=1\cdot10^{-28}gm/cm^{3}$ the magnitude $R/r_{g}=1.8$ and , therefore,
the acceleration at the time is negative However, at the density $\rho
=2\cdot10^{-28}gm/cm^{3}$ $R/r_{g}=0.9$ and the acceleration at the moment is
positive . It is equal to $1\cdot10^{-8}$ $cm/s^{2}$ which is half as large as
this magnitude resulting from the value of $q_{0}=-1$ that was found in \cite{Riess}

Fig. 4 shows the plots of the velocity and acceleration for $\bar{E}$ $>1$ .
In that case the acceleration is positive up to $R/r_{g}=4.49$. This magnitude
increases with the increase of $\bar{E}$.%

\begin{figure}
[h]
\begin{center}
\includegraphics[
height=1.9121in,
width=3.103in
]%
{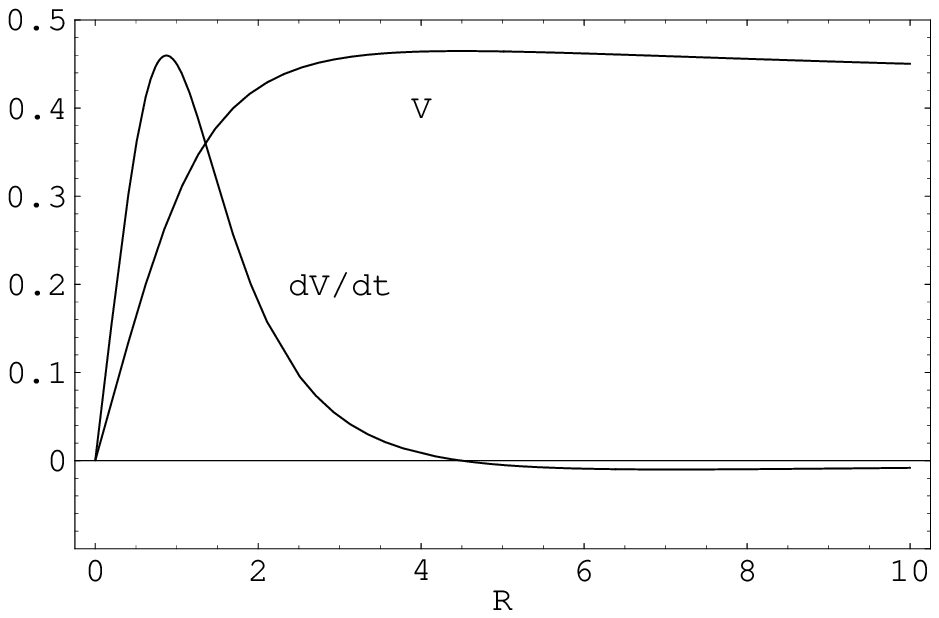}%
\end{center}
\end{figure}

\section{Conclusion}

The model does not yet gives the realistic description of the Universe
evolution scenario. We do not take into account the matter pressure. In spite
of this, the conclusions about "asymptotical" singularity and the simple
explanation of the expansion with acceleration worth of futher development.

\end{document}